\documentstyle[12pt]{article}

\begin{document}

\newcommand{\nc}{\newcommand}

\nc{\beq}{\begin{equation}}
\nc{\eeq}{\end{equation}}
\nc{\beqa}{\begin{eqnarray}} \nc{\eeqa}{\end{eqnarray}}
\nc{\lsim}{\begin{array}{c}\,\sim\vspace{-21pt}\\< \end{array}}
\nc{\gsim}{\begin{array}{c}\sim\vspace{-21pt}\\> \end{array}}
\nc{\el}{{\cal L}}
\nc{\D}{{\cal D}}
\def\boxit#1{\vbox{\hrule\hbox{\vrule\kern3pt
\vbox{\kern3pt#1\kern3pt}\kern3pt\vrule}\hrule}}
\def\smallbox{\boxit{\hbox to 1.5pt{
               \vrule height 1.0pt width 0pt \hfill}}}

\renewcommand{\thefootnote}{\fnsymbol{footnote}}
\setcounter{footnote}{0}
\begin{titlepage}

\def\thefootnote{\fnsymbol{footnote}}

\begin{center}

\hfill IASSNS-HEP-99/89\\
\hfill YCTP-P27-99\\
\hfill hep-th/9911029\\
\hfill November, 1999\\

\vskip .5in

{\Large \bf
Anomaly Mediation in Supergravity Theories
}

\vskip .45in

{\large
Jonathan A.\ Bagger,$^{a}$ Takeo Moroi,$^{b}$ and Erich Poppitz$^{c}$
}

\vskip 0.2in

{\em ${}^{a}$
  Department of Physics and Astronomy,
  The Johns Hopkins University\\
  Baltimore, MD 21218, USA}

\vskip 0.1in

{\em ${}^{b}$
  School of Natural Sciences,
  Institute for Advanced Study\\
  Princeton, NJ 08540, USA}

\vskip 0.1in

{\em ${}^{c}$
  Physics Department,
  Yale University\\
  New Haven, CT 06520-8120, USA}

\end{center}

\vskip .4in

\begin{abstract}

We consider the effects of anomalies on the supersymmetry-breaking
parameters in supergravity theories.  We construct a supersymmetric
expression for the anomaly-induced terms in the 1PI effective
action; we use this result to compute
the complete one-loop formula for the anomaly-induced gaugino
mass.  The mass receives contributions from the super-Weyl,
K\"ahler, and sigma-model anomalies of the supergravity
theory.  We point out that the
anomaly-mediated gaugino mass can be affected by local counterterms
that cancel the super-Weyl-K\"ahler anomaly.  This implies that
the gaugino mass cannot be predicted unless the full high-energy
theory is known.

\end{abstract}
\end{titlepage}

\renewcommand{\thepage}{\arabic{page}}
\setcounter{page}{1}
\renewcommand{\thefootnote}{\#\arabic{footnote}}
\setcounter{footnote}{0}

\section{Introduction and summary}

Supersymmetry at the electroweak scale is an appealing possibility
for physics beyond the standard model.  Not only does it render
the ratio of the electroweak scale to the Planck scale, $M_W/M_
{\rm P} \sim 10^{-17}$, stable against quadratically divergent
radiative corrections, but it can also explain the size of the
hierarchy if supersymmetry is broken by some nonperturbative
mechanism.

It is, therefore, an urgent and open problem to find a viable
and testable method for dynamically breaking supersymmetry
and generating weak-scale masses for the superpartners
of the known elementary particles.  The problem has two
aspects, which are not necessarily distinct.  First, one
has to dynamically break supersymmetry, and second, one
has to communicate the supersymmetry breaking to the
visible-sector particles.  The main ideas for this are
supergravity and gauge mediation.

In this paper we will consider
supergravity-mediated supersymmetry breaking.  In such
theories, the communication of supersymmetry breaking
to the scalar fields is almost automatic.  For a generic
K\"ahler potential, once supersymmetry is broken in a hidden
sector, visible-sector scalars gain mass as a consequence
of the tree-level supergravity equations of motion.  The
scale of the mass, $M_0$, is set by dimensional analysis:
$M_0 \sim M_{\rm SUSY}^2/M_{\rm P} \sim m_{3/2}$, where
$M_{\rm SUSY}$ is the scale of supersymmetry breaking and
$m_{3/2}$ is the gravitino mass.  If $M_{\rm SUSY}$ is of
order an intermediate scale, $M_{\rm SUSY} \sim 10^{10}$
GeV, the soft scalar masses are of order $M_W$.

Gaugino masses, as well as $A$-terms, are more
difficult to obtain because they break a $U(1)_R$ symmetry.
The simplest way to generate these terms is to include
an effective gauge-singlet chiral superfield in the
low-energy effective Lagrangian.  The problem
is that such singlets are not necessarily
present in generic models of supersymmetry breaking.
Moreover, fundamental singlets can cause cosmological
difficulties, or destabilize the gauge hierarchy
through quadratically divergent radiative corrections.
It is important, therefore, to fully investigate
alternative mechanisms for generating gaugino masses.

It was recently pointed out by Randall and Sundrum~\cite{RS},
and Giudice, Luty, Murayama, and Rattazzi~\cite{GLMR}, that
once supersymmetry is broken in the hidden sector of a theory,
gaugino masses and $A$-terms are generated at one loop from
anomaly-related graphs.  This is an important result because
it shows that these terms are automatically present,
even in theories {\it without} gauge singlets.  The simplest
way to understand their result is to consider the one-loop
renormalization of the visible-sector gauge couplings,\footnote
{In this paper, we omit the index for the adjoint representation
for the gauge multiplets.  For non-Abelian gauge groups,
$W^\alpha W_\alpha$ and $F_{mn}F^{mn}$ should be understood
as $W^{(a)\alpha} W_\alpha^{(a)}$ and $F_{mn}^{(a)} F^{(a)mn}$,
where $(a)$ is the index of the adjoint
representation.  We use the conventions of Ref.~\cite{WB}.}
\footnote{
For simplicity, we do not include gauge kinetic functions.
In general, gauge kinetic functions can occur at the tree
or loop level.  If present, they can give non-anomalous
contributions to gaugino masses.  We are interested in
anomaly-induced gaugino masses, so we do not consider
them here.
}
\begin{equation}
\label{rsglmr}
{1 \over 4} \int d^2 \theta \left( 1\ -\ {g^2 b_0
\over 16 \pi^2} \log {\,\Lambda^2 \over \smallbox} \right) W^\alpha
W_\alpha \ +\ {\rm h.c.}~,
\end{equation}
where $b_0$ is the first coefficient of the beta function
and $\Lambda$ is the ultraviolet cutoff.  As pointed out in
Refs.~\cite{RS,GLMR}, the leading supersymmetry-breaking effect
can be found, in the absence of Planck-scale hidden-sector
expectation values, by replacing the ultraviolet cutoff in
(\ref{rsglmr}) with a supersymmetry-breaking spurion
superfield,
\begin{equation}
\label{replacecutoff}
\Lambda\ \rightarrow\ \Lambda\, \exp(m_{3/2}\, \theta^2)~.
\end{equation}
This leads, after substitution in (\ref{rsglmr}), to the
following ``anomaly-induced" gaugino mass,
\begin{equation}
\label{mass}
m_{1/2}\ =\ - {g^2 b_0 \over 16 \pi^2}\, m_{3/2} ~.
\end{equation}
One-loop $A$-terms are induced in a similar way.  This mechanism
for generating gaugino masses gives rise to an exciting new
phenomenology \cite{pheno}.

In what follows we will take a closer look at the physics that
underlies anomaly mediation, in the full supergravity context.
Our starting point will be the general supergravity Lagrangian.
The classical symmetries of this Lagrangian are super-Weyl-K\"ahler
invariance (or, in components, simply K\"ahler invariance), as
well as various global symmetries that act as sigma-model
isometries.  These symmetries induce anomalous chiral rotations
on the fermions.  We shall see that anomaly mediation results
from quantum anomalies in these classical symmetries.

In Section 2 we will elaborate on these points, and present
the terms in the 1PI effective action that are induced by
the chiral anomalies associated with the super-Weyl-K\"ahler
and sigma-model symmetries.  These nonlocal terms are entirely
determined by Bose symmetry, unitarity, analyticity, and gauge
invariance.  They are, therefore, independent of the choice of
regulator\footnote
{A 1PI argument for the regulator independence of anomaly
mediation, based on local supersymmetry, was also given
in Ref.~\cite{GLMR}.}
\cite{DZ}.
By demanding local supersymmetry, we will find the unique
superspace extension of these nonlocal anomaly-induced terms
\cite{CO}.

In Section 3 we will discuss the conditions under which these
nonlocal terms are present in the effective action.  We will
argue that they are indeed present in the case of interest.
We will then derive the general formula for the one-loop
anomaly-induced contribution to the gaugino mass, valid for
hidden-sector models with arbitrary expectation values,
\begin{equation}
 m_{1/2}\ =\ - {  g^2 \over 16 \pi^2}  \left[\, (3 T_G - T_R)\, m_{3/2} +
 (T_G - T_R) \, K_i F^i + {2\, T_R \over d_R}\,\, (
 {\log \det}\, K{\vert_R}^{\prime \prime} )_{, i}
\,F^i \,\right]~.
\end{equation}
In this expression, $3 T_G - T_R$ is the beta function, $b_0$,
where $T_G$ is the
Dynkin index of the adjoint representation, normalized to
$N$ for $SU(N)$, and $T_R$ is the Dynkin index associated
with the representation $R$ of dimension $d_R$, normalized
to 1/2 for the $SU(N)$ fundamental.  A sum over all matter
representations is understood.  The second and third terms
in this expression involve the K\"ahler potential, $K$,
and the expectation values of the auxiliary fields, $F^i$,
evaluated in the Einstein frame.  This result for $m_{1/2}$
generalizes the one given in Refs.~\cite{RS,GLMR}, and
reduces to it when all expectation values are much less
than $M_{\rm P}$.

In Section 4 we will see that anomaly-induced gaugino
masses can be modified by contributions from Planck-scale physics.
In general, we expect such contributions because supergravity
is a nonrenormalizable, effective field theory, valid for
energies much less than $M_{\rm P}$.  In string theory,
for example, the heavy string modes can
induce anomalous Green-Schwarz-like terms in the effective action.
We shall see that these terms can contribute
to the gaugino masses.   This illustrates that
the gaugino masses can
be significantly modified -- or canceled altogether -- by
high energy contributions that are outside the realm of the
effective supergravity theory.

\section{Anomalies in matter-coupled supergravity}

In this section, we will define the classical symmetries of
matter-coupled supergravity:  super-Weyl-K\"ahler invariance
and sigma-model isometries.  We will compute their quantum
anomalies and write down, following Cardoso and Ovrut \cite{CO},
the terms they induce in the 1PI effective action.  We will
restrict our attention to the mixed super-Weyl-K\"ahler gauge
and sigma-model gauge anomalies because they are the
anomalies that are relevant for gaugino masses.

The super-Weyl-K\"ahler and sigma-model symmetries of classical
supergravity Lagrangian can be given a full superspace description
(see \cite{WB}).  However, for the case at hand, it is perhaps
simpler to study them through their action on the fermions in
the supergravity Lagrangian.  To this end, let us consider the
kinetic terms for the fermions:
\begin{eqnarray}
 \label{fermionkineticterms}
 \el_{\rm kin} &=&
 - i K_{ij^*}\, \bar\chi^j \bar\sigma^a  \left[ \left\{
 \partial_a + {i \over 6}\, b_a - {1\over 6}\, (K_j \partial_a A^j -
 K_{j^*} \partial_a A^{*j} )  \right\}
 \chi^i  + \Gamma^i_{j k}\partial_a A^j
 \chi^k \right]\nonumber \\
  & & - i \bar\lambda \bar\sigma^a \left( \partial_a - {i \over 2}
 b_a \right) \lambda~,
\end{eqnarray}
where $A^i$ and $\chi^i$ are the scalar and fermionic components
of $i$-th chiral superfield, the $\lambda$ are the gauginos, and
$b_a$ is the auxiliary vector field in the supergravity multiplet.
Furthermore, let us define $K_i \equiv \partial K/\partial A^i$,
and $K_{ij^*}$ and $\Gamma^i_{jk}$ to be the K\"ahler metric and
K\"ahler connection, respectively,
\begin{equation}
K_{ij^*}\ =\ \frac{\partial^2 K}{\partial A^i \partial A^{*j}}~,
\qquad\qquad
K_{\ell j^*} \Gamma^\ell_{ik}\ =
\ \frac{\partial K_{ij^*}}{\partial A^k} \ .
\end{equation}
We choose to write Eq.~(\ref{fermionkineticterms}) without
eliminating the auxiliary field $b_a$ because we wish to
preserve off-shell supersymmetry.  We assume that the gauge
fields couple in the
usual manner.

The fermion kinetic terms (\ref{fermionkineticterms}) contain
connections for three local symmetries. The first is a $U(1)_R$
symmetry that is part of the superconformal group: it acts on
the gauginos, the matter fermions, and the
supergravity auxiliary field as follows,
\begin{eqnarray}
\label{u1rsc}
\lambda &\rightarrow& e^{i \alpha/2}\, \lambda  \nonumber \\
\chi &\rightarrow& e^{- i \alpha/6}\, \chi \\
b_a &\rightarrow& b_a\, +\, \partial_a \alpha~. \nonumber
\end{eqnarray}
The connection $b_a$ shifts under the $U(1)_R$ symmetry because it
is the gauge field of superconformal supergravity.  (Recall
that the action and auxiliary field structure of $N=1$ Einstein
supergravity can be obtained by gauge fixing the superconformal
supergravity action; see e.g.~\cite{vN}.)  From these transformations
we see that the chiral anomaly associated with the superconformal
symmetry is proportional to $b_0 = 3 T_G - T_R$.

The second symmetry is a local $U(1)_K$ symmetry that acts on
the matter fermions but not the gauginos.  It is compensated
by a shift of the K\"ahler connection:
\begin{eqnarray}
\label{u1prime}
\chi &\rightarrow & e^{-i \beta/6}\, \chi \\
i(K_m \partial_a A^m - K_{m^*} \partial_a A^{*m}) &\rightarrow  &
i(K_m \partial_a A^m - K_{m^*} \partial_a A^{*m})\, +\,
\partial_a \beta \nonumber ~.
\end{eqnarray}
The chiral anomaly of this symmetry is proportional to $T_R$.
Note that the $U(1)_R \times U(1)_K$ symmetry of the fermion
kinetic terms is explicitly broken by other terms in the
supergravity action.  The unbroken symmetry
is super-Weyl-K\"ahler symmetry, which in component form is
usually called K\"ahler invariance \cite{WB}.
This, however, does not affect the nonlocal anomaly terms that
are induced in the 1PI effective action.

The third connection entering the fermion kinetic terms is
the sigma-model connection, $\Gamma^i_{jk} \equiv K^{i \ell^*}
\partial_j K_{k \ell^*}$, where $K^{i \ell^*}$ is the inverse
K\"ahler metric.  It acts as a $U(N)$ connection on the
matter fermions, where $N$ is the number of chiral multiplets
in the theory.  However, only the $U(1)$ subgroup contributes
to the gaugino masses.  Therefore, in what follows, we
restrict our attention to this $U(1)$, and write
\begin{equation}
\label{sigmamodelconnection}
\Gamma^i_{jk} \partial_a A^j\ =
\ d_R^{-1}\,\delta^i_k \Gamma_{j \ell}^{\ell}
\partial_a A^j  \ + \ \cdots
\ =\ d_R^{-1}\,\delta^i_k \, ({\log \det}\, K{\vert_R}^{\prime
\prime})_{,j} \partial_a A^j\ +\ \cdots~,
\end{equation}
where the dots denote the non-singlet part of the sigma-model
connection, and $K{\vert_R}^{\prime\prime}$ is the K\"ahler
metric restricted to the representation $R$. In
Eq.~(\ref{sigmamodelconnection}), a sum over all matter
representations $R$ is understood, and the sum over $\ell$
is restricted to the appropriate representations.

The super-Weyl, K\"ahler and sigma model symmetries are
all anomalous.  The connections $b_a$, $K_m\partial_a A^m
- K_{m^*} \partial_a A^{*m}$, and $\Gamma^i_{jk} \partial_a A^j$
can be viewed as background gauge fields coupled to
anomalous currents.  Consider, for example, the connection $b_a$.
The gauge triangle diagram gives rise to a nonlocal term
in the 1PI effective action, whose form is determined by Bose
symmetry, unitarity, analyticity, and gauge invariance
\cite{DZ}:
\begin{equation}
\label{componentanomaly}
\Delta \el\ = \ {g^2 \over 96 \pi^2}\, (3 T_G - T_R)\,
{\partial_a b^a \over \smallbox }\, F_{mn} \tilde{F}^{mn}~.
\end{equation}
In this expression, $F_{mn}$ and $\tilde{F}^{mn}$ are the
field strength and the dual field strength associated with
a set of dynamical gauge fields, and $b_a$ is the background
gauge field that couples to the anomalous $U(1)_R$ current.
Under a background gauge transformation, $b_a \rightarrow
b_a + \partial_a \alpha$, $\Delta \el$ has a local variation
that expresses the $U(1)_R$ anomaly.  Similar terms appear
for the other two currents.

Note that the anomaly (\ref{componentanomaly}) depends on the
quadratic Casimirs $T_G$ and $T_R$.  These Casimirs include
only those fermions that are effectively massless at the
momenta of interest.  If a fermion propagating in the loop
has a Dirac mass $m$, it decouples from the anomaly for
external momenta $q^2 \ll m^2$.  (The situation for masses
generated by spontaneous symmetry breaking will be discussed
in Section 4.)

In a supersymmetric theory, the above anomalies can be lifted
to superspace \cite{WB}.  Following Cardoso and Ovrut \cite{CO},
the result is as follows
\begin{eqnarray}
\Delta \el &=& - {g^2 \over 256\pi^2} \int d^2 \Theta
{}~2{\cal{E}}~W^\alpha W_\alpha
\ {1 \over \smallbox} \left( {\bar\D}^2 - 8R
\right) \nonumber \\ &&
\times
\left[4 (T_R - 3 T_G)\, R^+
- {1 \over 3} T_R \, \D^2 K +
\frac{T_R}{d_R} \D^2 \,\log \det K {\vert_R}^{\prime \prime}
\right]\,+\, {\rm h.c.}
\label{ovrutterm}
\end{eqnarray}
The first term, which contains the $R^+$ superfield,
arises from the $U(1)_R$ anomaly.  It is proportional
to the beta function, $b_0 = 3 T_G - T_R$.  The second
and third terms express the $U(1)_K$ anomaly and the
$U(1)$ piece of the sigma-model anomaly.  Together they
have the appropriate local variations to produce the
quantum anomalies associated with the super-Weyl-K\"ahler
and $U(1)$ sigma-model symmetries \cite{CO}.

The anomaly for each connection can be obtained by a background
superfield calculation of the two-point gauge superfield
Green's function, with a single insertion of the appropriate
background field (the superspace curvature $R$ for the first
term, and $K$ or $\log\det K^{\prime\prime}$ for the last two
terms).  Therefore this result should be interpreted as the
leading contribution in a background-field expansion.  In
particular, the inverse Laplacian should be evaluated in
flat spacetime.

To see that the superspace expression (\ref{ovrutterm})
does indeed reproduce the correct component anomaly terms,
one must expand in components.  Consider, for example,
the $U(1)_R$ term.  The component expansion of the superfield
$R^+$ contains the following terms \cite{WB},
\begin{equation}
\label{rdagger}
R^+\ =\  - {1 \over 6} \left[ M^* + \cdots + \bar\Theta^2
\left( - {1\over 2} {\cal R} + i e^m_a
{\cal D}_m b^a  + \cdots \right) \right]~,
\end{equation}
where ${\cal R}$ is the Einstein curvature scalar.
If we substitute (\ref{rdagger}) into the first
term in (\ref{ovrutterm}), and take the $\Theta^2$ component
of $W^\alpha W_\alpha$ and the lowest component of $(\bar\D^2
- 8 R) R^+$, we recover the same nonlocal $U(1)_R$ anomaly
as in (\ref{componentanomaly}).

The superspace expression (\ref{ovrutterm}) also contains
the supersymmetric partners of the $U(1)_R$ anomaly.  In
particular, it contains the nonlocal term
\begin{equation}
\Delta {\cal L}\ =\ \frac{g^2}{192\pi^2}\, (3T_G-T_R)\,
\frac{{\cal R}}{\smallbox}\, F_{mn} F^{mn}~.
\label{conformal}
\end{equation}
This term arises from a one loop graph with one graviton
and two gauge boson vertices.  Under a conformal rescaling
of the metric, $g_{mn} \rightarrow \exp(2 \lambda) g_{mn}$,
the curvature scalar shifts as ${\cal R} \rightarrow {\cal R}
+ 6\,\smallbox\lambda + {\cal O}(\lambda^2)$.  We see that
the nonlocal term (\ref{conformal}) also expresses the
conformal anomaly of the theory.

In the following section, we will compute the full
anomaly-mediated gaugino mass term.  Before we do
that, however, we first discuss the conditions
under which Eq.~(\ref{ovrutterm}) is valid.  We start
by noting that our derivation required two essential
ingredients:
\begin{itemize}
\item[$(i)$] {A nonlocal anomaly term of the type given
in (\ref{componentanomaly}).}
\item[$(ii)$] {Local supersymmetry of the 1PI action.}
\end{itemize}

The first item is obvious.  If there are no anomalies, there
are no anomaly-mediated contributions to gaugino masses.   For
visible-sector gaugino masses, the associated anomaly diagrams
contain visible-sector fermions in the loop.  The anomaly
receives contributions from all fermions whose mass is
less than the weak scale, $M_W$.

The second item says that the superspace expression only
holds when the superpartners of the loop fermions are
active at the scale of interest.  For gaugino masses,
the loop contains visible-sector particles.  Since all
visible-sector superparticles have weak-scale masses, the
effective theory is essentially supersymmetric.
This implies that the expression (\ref{ovrutterm}) can
indeed be used to extract gaugino masses.

A more refined analysis confirms this intuition.
Consider the case of a visible sector in which there
are scalar mass terms present at the tree level, but no
tree-level fermion masses.  Since the fermions are massless,
triangle diagrams induce anomalies in the anomalous currents.

Is the Cardoso-Ovrut result valid for this case?
Eq.~(\ref{ovrutterm})
depends on graphs that contain just one background-field
insertion in the gauge two-point function.  Since the scalar
mass terms arise from such insertions, one should, in fact,
sum over all hidden-sector insertions.  These insertions
have important effects.  For example, they can change
the conformal anomaly, Eq.~(\ref{conformal}).  If the
scalars have mass, their contribution must decouple at
momenta below their mass.  The decoupling of the scalars
is not described by the Cardoso-Ovrut term because the
extra insertions are not included.

Fortunately, the scalar mass insertions have no effect on
the diagrams that give rise to gaugino masses. One way
to see this is to note that gaugino masses arise as a result of
loops with Pauli-Villars regulator fields. Therefore the
anomaly-induced gaugino masses are present whether or not
the visible-sector scalar fields have tree-level masses.

By way of contrast, consider now the case of the hidden-sector
loops that contribute to the hidden-sector gauge two-point
function.  Naively, one might think that these loops would
generate visible-sector scalar masses through the Cardoso-Ovrut
term.  However, they do not because conditions $(i)$ and $(ii)$
both fail in the hidden sector.  In particular, hidden-sector
fermions are generally not massless, so the terms of the form
(\ref{componentanomaly}) are not present at scales below $M_{\rm SUSY}$.
Second, even if there are hidden-sector massless fermions,
their superpartners are heavy, typically of order $M_{\rm SUSY}$.
Therefore the Cardoso-Ovrut
term cannot be used to infer one-loop anomaly-induced masses
for the visible-sector supersymmetric scalars.

\section{Anomaly-induced gaugino masses}

In the previous section, we found that the super-Weyl-K\"ahler
and sigma-model anomalies generate nonlocal terms in the 1PI
effective action.  The terms are induced by vector-field
two-point diagrams that contain one additional insertion of
a supergravity or hidden-sector background field.  In this
section we will see that they give
rise to gaugino masses when the background fields obtain
supersymmetry-breaking expectation values.

The anomaly-induced contribution to the gaugino mass can be
easily computed from (\ref{ovrutterm}).  One first extracts
the $\Theta^2$ components of the terms multiplying $W^\alpha
W_\alpha$.  This gives
\begin{equation}
\label{m1/2}
{ g^2 \over 16 \pi^2} \left[ {M^* \over 3}\, (3 T_G - T_R)
\ +\ \frac{2}{3}\, T_R \, K_i F^i \ -
\ \frac{2T_R}{d_R}\, (\log \det K{\vert_R}^{\prime  \prime}
)_{, i} F^i \right]~.
\end{equation}
In this expression, the expectation values of the auxiliary
fields $M$ and $F^i$ are evaluated in the general
``supergravity frame," in which the Einstein term is of the
form $\exp(-K/3) {\cal R}$.

To make contact with the usual supergravity Lagrangian
(see e.g.~\cite{WB}), one must transform to the
``Einstein frame," in which the Einstein term
takes its canonical form.  The frame transformation is
accomplished by a Weyl rescaling of the metric
and a redefinition of the fermionic and auxiliary fields.  The
relevant transformations are as follows \cite{WB},
\begin{eqnarray}
\label{einsteinframetransforms}
e_c{}^{m} &=& e^{- 2 \sigma}\, e^\prime{}_c{}^{m} \nonumber \\
\lambda &=& e^{- 3 \sigma}\, \lambda^\prime \nonumber \\
M^* &=& e^{- 2 \sigma}\, ( M^{\prime\, *} - F^{\prime\, i}  K_i ) \\
F^i &=& e^{- 2 \sigma}\, F^{\prime\, i} \nonumber ~,
\end{eqnarray}
where the primed quantities are in the Einstein frame
and $\sigma = K/12$.  Note that the field $M$ transforms
inhomogeneously, while the $F^i$ transforms homogeneously.

Using these results, it is not hard to write the anomaly-induced
gaugino mass in the usual Einstein frame.   One first
redefines the auxiliary fields as in (\ref{einsteinframetransforms})
and then one drops the primes.  This gives the complete
one-loop anomaly-induced contribution to the gaugino mass,
\begin{equation}
m_{1/2}\ =\ - {g^2 \over 16 \pi^2}
\,\left[(3 T_G - T_R)\, m_{3/2}\  +\ (T_G - T_R)\, K_i F^i
+ \frac{2T_R}{d_R}\,
\left( \log \det K {\vert_R}^{\prime \prime} \right)_i F^i
\right]~.
\label{gauginomasseinstein}
\end{equation}
In this expression, we have substituted the Einstein-frame
expectation value of $M$,
\begin{equation}
M^*\ =\ -3 e^{K/2}\, P^*\ \equiv\ -3 m_{3/2}~,
\end{equation}
where $P$ is the superpotential.  We also require the field
$F^i$ to be evaluated in the Einstein frame, where
\begin{equation}
F^i\ =\ - e^{K/2}\, K^{i j^*}\, ( P^*_{j^*} + K_{j^*} P^*)~.
\end{equation}

Let us now consider several limits of (\ref{gauginomasseinstein}).
The first is when there are no Planck-scale expectation values
in the hidden sector.  In that case, it is easy to check
that Eq.~(\ref{gauginomasseinstein}) reduces to the result given in
Ref.~\cite{RS, GLMR},
\begin{equation}
m_{1/2}\  =\ - \frac{g^2}{16\pi^2}\, (3 T_G - T_R)\, m_{3/2}~,
\end{equation}
with $b_0 = 3 T_G- T_R$.

The second limit is when the K\"ahler potential is of the
``sequestered-sector'' form \cite{RS},
\begin{equation}
\label{sequestered}
K\ =\ -3\log \left[
\, -\frac{1}{3} Q^+ Q \ +\ f(H^+, H)\,\right]\ .
\label{K_RS}
\end{equation}
Here $Q$ and $H$ denote observable and hidden-sector
chiral superfields, respectively, and $f$ is a real function
with $\langle f\rangle =1$.  This K\"ahler potential has
the property that it generates no tree-level soft scalar
masses for the visible-sector fields.  It also obeys the
relation
\begin{equation}
\label{properties}
\frac{1}{d_R} \,\log\det K{\vert_R}^{\prime\prime}
\ = \ { 1\over 3}\,K
\end{equation}
for vanishing expectation values of the observable fields.
(Recall that the derivatives in $K^{\prime\prime}$ are taken
with respect to visible fields, and the metric is projected
on the representation $R$ of dimension $d_R$ of the relevant
gauge group.)  This implies that the
contributions from the K\"ahler and sigma-model connections
cancel, leaving \cite{PR}:
\begin{equation}
\label{m1/2sequestered}
m_{1/2}\  = \ - {g^2 \over 16 \pi^2}
\,(3 T_G - T_R)\, \left( m_{3/2} +
{1 \over 3} K_i F^i \right)~.
\end{equation}
In the absence of Planck-scale expectation values, this
reduces to the
result given in Refs.~\cite{RS,GLMR}.

To highlight the importance of the first term in
(\ref{gauginomasseinstein}), it is useful to consider
a third example, that of a ``no-scale'' model, with
K\"ahler potential
\begin{equation}
\label{noscale}
K\ =\ - 3 \log \left( T + T^+
- \frac{1}{3} Q^+ Q
- \frac{1}{3} H^+ H
\right)~.
\end{equation}
This K\" ahler potential is of the sequestered-sector
type (if we treat the modulus $T$ as a hidden-sector field),
so the gaugino mass is given by Eq.~(\ref{m1/2sequestered}).
If the supersymmetry breaking is dominated by the $F$
component of the modulus $T$, and the cosmological
constant is canceled by adding a constant to the
superpotential, it is easy to see that $K_T F^T =
- 3m_{3/2}$.  We see that the anomaly-induced gaugino mass
vanishes in this model.

Let us now compare our results with previous work on
supergravity anomalies \cite{DFKZ, IL}.  These
papers contain a nonlocal anomaly-induced term that
is similar to our expression (\ref{ovrutterm}), except
for the fact that the first term
is missing, and the second term is
proportional to $(T_G - T_R)\,D^2 K$.  The expression
in these papers has the correct anomalous variation
under super-Weyl-K\"ahler transformations, but it
does not transform correctly under super-Weyl or
K\"ahler alone.  This can be traced to the missing
$R^+$ term in their expression for the anomaly.
The expressions in these papers completely
miss the super-Weyl contribution to the gaugino mass,
proportional to $(3 T_G - T_R) m_{3/2}$.  (This point
was also made in \cite{MK}.)  Note that in the
``no-scale'' model considered above,
the $R^+$ contribution precisely cancels the terms
that involve the K\"ahler potential.

The nonlocal terms induced by the super-Weyl anomaly are
important for a different reason as well.  As will be shown
in \cite{BMP2}, the transformation to the Einstein frame
is actually a super-Weyl field redefinition.  This
redefinition gives rise to an anomalous Jacobian
in the Einstein-frame Lagrangian.  The Jacobian can
be obtained by a
shift in  (\ref{ovrutterm}).
The Jacobian ensures that supersymmetry transformations
are not anomalous in the Einstein frame.  The issues of
frame dependence and the super-Weyl anomaly will be
discussed in a separate publication \cite{BMP2}.

\section{Counterterms from high energies}

In this section we will discuss high-energy
contributions to the effective Lagrangian and their
possible effects on anomaly-mediated gaugino masses.
These contributions can be generated by heavy modes
that are not included in the low-energy supergravity
theory.

We first consider contributions that are induced
by physics below the Planck scale \cite{PR}.
Consider, for example,
a theory that contains a visible-sector chiral multiplet
with a mass of order $M$, where $M$ is much
larger than the scale of the effective
field theory.  Let us assume that
the multiplet gets its mass from symmetry breaking,
through a superpotential term of the form
\begin{equation}
{\cal L}\ =\ \int d^2\Theta \ {\cal E}\,\Phi \,Q^2
\,+\, {\rm h.c.}~,
\end{equation}
where $Q$ is in a real representation
of the gauge group.
In this expression, $\Phi$ is a spurion superfield with
expectation value $\langle \Phi \rangle = \Phi_0 +
F \Theta^2$.  Let us include this multiplet
in the low-energy theory, and then decouple it by
taking $\Phi_0 \rightarrow M$.  The decoupling
is easily done using a field redefinition,
$Q \rightarrow Q/\sqrt{\Phi/\Phi_0}$,
which rotates away the supersymmetry-breaking
expectation value, $F \Theta^2$.
(This is similar to the D'Hoker-Farhi method for integrating
out a heavy top quark \cite{dhoker-farhi}.  See also
\cite{PR, KSS}.)
The field redefinition is, of course, anomalous;
its Jacobian can be computed from the $\log \det K
{\vert_R}^{\prime \prime}$ part of (\ref{ovrutterm}):
\begin{equation}
{\Delta \cal L}\ =\ {g^2 \over 32\pi^2} \,T_R\,
\int d^2\Theta \ 2{\cal E}\,\log(\Phi/\Phi_0)
 \, W^\alpha W_\alpha\,+\, {\rm h.c.}
\end{equation}
As $M \rightarrow \infty$, the Jacobian induces an
additional contribution to the gaugino mass,
\begin{equation}
\Delta m_{1/2}\ =\ {g^2\over 16 \pi^2}\,T_R\,{F\over M}\ .
\end{equation}

This example shows that anomaly-induced gaugino masses
can depend on physics beyond the weak scale.  In fact,
they can also depend on physics beyond the Planck scale.
To see this, consider the case of superstring
theory, which reduces to supergravity for
energies below the Planck scale.  If we restrict ourselves
to the low-energy supergravity theory, there is no reason
for the super-Weyl-K\"ahler transformations to be
anomaly-free.  (The low-energy supergravity theory is
well defined whether or not the transformations are
anomalous.)  In string theory, however, it is sometimes
necessary to cancel the anomalies.

The low-energy supergravity that arises from superstring
compactification typically depends on the size and shape
of the extra compact space-time dimensions.  In supergravity
theory, this information is encoded in the moduli, which
are most easily described in terms of sigma models.
These sigma models can have various nonlinear
symmetries, such as target space duality, which are
reflections of exact quantum symmetries of the underlying
string theory.   These nonlinear symmetries can act on the
supergravity theory as super-Weyl-K\"ahler transformations,
in which case the relevant super-Weyl-K\"ahler anomalies
must be canceled.

As a simple example, consider the $SU(1,1)/U(1)$ supersymmetric
sigma model, which describes the moduli dynamics in certain
toroidal superstring compactifications (see e.g. \cite{polchinski}).
The K\"ahler potential for the (complex) modulus $T$ is given by
\begin{equation}
\label{kahlercp1}
K \ =\ -\log \left(T + T^+  \right)~,
\end{equation}
in the usual units with $M_P = 1$.
(We neglect matter-field contributions to $K$.)
Target-space duality acts on $T$ through
an $SL(2,R)/Z_2$ transformation.
Under $T \rightarrow 1/T$, for example, the K\"ahler potential
changes as follows,
\begin{equation}
\label{kahlercp2}
K(1/T, 1/T^+) \ =\ K(T, T^+)\ +\ \log\, T\ +\ \log\, T^+~.
\end{equation}
This induces a super-Weyl-K\"ahler transformation
in the low-energy supergravity theory.  Now, if duality is an
exact quantum symmetry, the corresponding super-Weyl-K\"ahler
anomaly must be canceled by other terms in the low-energy
effective action.  These terms are local because they arise
from integrating out high-energy modes; their variation
under the super-Weyl-K\"ahler transformation precisely cancels
the anomaly.

The question for us here is whether the anomaly-canceling
counterterms contribute to the gaugino masses.  In general,
they do, by an amount that depends on unknown
high energy physics.  Consider, for example, the following
term:\footnote{
The $\log \eta(iT)$ can be viewed as a contribution to a
gauge kinetic function $f(T)$.}
\begin{equation}
\label{wzterm}
\Delta {\cal L}\ =
\ {g^2 \over 16 \pi^2}\,(T_G - T_R)\, \int d^2 \Theta ~2 {\cal{E}}~
\log \eta(iT)\ W^\alpha W_\alpha\, +\, {\rm h.c.}~,
\end{equation}
where $\eta(iT)$ is the Dedekind eta function.  This
term shifts under $T \rightarrow 1/T$, and cancels
the super-Weyl-K\"ahler
anomaly arising from (\ref{kahlercp2}).\footnote{The
coefficient in $\Delta {\cal L}$ can be different
when visible-sector matter fields are included.}
If $T$ has a Planck-scale expectation value, $\langle T
\rangle = T_0 + F \Theta^2$, Eq.~(\ref{wzterm})
gives the following contribution to the gaugino mass
\begin{equation}
\Delta m_{1/2}\ =\ -{g^2 \over 96 \pi}\,(T_G - T_R)\,
E_2(iT_0) \,F\ ,
\label{eisen}
\end{equation}
where $E_2(iT)$ is the second Eisenstein series.

Other anomaly-canceling terms are also possible.  A second
example is provided by the Green-Schwarz term from \cite{DFKZ},
\begin{equation}
\label{derendinger}
\Delta {\cal L}\ =
\ {g^2 \over 32 \pi^2}\,(T_G - T_R)\, \int d^2
\Theta ~2 {\cal{E}} ~\left( \bar{\D}^2 - 8 R \right)
\,K(T^+,T)\, \left(\Omega - L\right)\, +\, {\rm h.c.}
\end{equation}
Here $K$ is the K\"ahler potential, $\Omega$ is the Chern-Simons
superfield, and $L$ is a linear multiplet that is required for
gauge invariance (for details we refer the reader to \cite{DFKZ}).
This term shifts the gaugino mass by
\begin{equation}
\Delta m_{1/2}\ =\
{g^2 \over 16 \pi^2}\,(T_G - T_R)\,K_T F\ ,
\end{equation}
which is clearly not the same as (\ref{eisen}).

These examples show that the gaugino masses predicted by
anomaly meditation are affected by unknown high-energy
physics.  On the one hand, this makes it difficult to experimentally
test the ideas behind anomaly mediation.  On the other, the
dependence on string-scale physics opens a window to physics
at the highest energies.  It is an interesting and important
problem to study the anomaly-canceling terms in various
string vacua.

A final example that illustrates the dependence on high energy
physics is the ambiguous separation of the ``K\"ahler potential''
and the ``superpotential'' in the bare Lagrangian.  At the tree
level, the following two superspace Lagrangians give rise to the
same component Lagrangians:
\begin{eqnarray}
{\cal L}_0 &=&
\int d^2\Theta ~2{\cal E}~\left[ \frac{3}{8}
\left( {\bar\D}^2 - 8 R \right) ~\exp(-K/3) + P \right]\ +\ \cdots,
\label{L_WB}
\\
{\cal L}'_0 &=&
\int d^2\Theta ~2{\cal E}~\left[ \frac{3}{8}
\left( {\bar\D}^2 - 8 R \right) ~\exp[-(K+\log P + \log P^+)/3] + 1 \right]
 \ +\ \cdots.
\label{L'}
\end{eqnarray}
In these expressions, we have suppressed the gauge-field-dependent
terms.  At tree-level, Eqs.~(\ref{L_WB}) and (\ref{L'})
are related by a super-Weyl
transformation.  Beyond tree-level, the equivalence does not persist
because of the super-Weyl-K\"ahler anomaly.  The one-loop
effective actions derived
from these Lagrangians are related as follows
\begin{equation}
{\cal L}'\ =\ {\cal L}
\ -\ \left[ \frac{g^2}{32\pi^2}\, (T_G - T_R)\,
\int d^2\Theta ~2{\cal E}~\log P ~W^\alpha W_\alpha\ +\ {\rm h.c.}
\right].
\end{equation}
This relation is easily derived by writing $R' =
R - {\bar{\cal D}}^2 \log P^+/24$ in (\ref{ovrutterm})
and then redefining the K\"ahler potential, $K'
= K + \log P + \log P^+$.

Our formula, given in Eq.~(\ref{gauginomasseinstein}), was
derived using the Lagrangian (\ref{L_WB}).  If we had used
the Lagrangian (\ref{L'}), we would have found the gaugino
mass shifted by\footnote
{In the sequestered-sector scenario~\cite{RS}, this ambiguity does
not arise.  In this scenario, there is no direct coupling between
hidden- and observable-sector fields in the bare Lagrangian.  Therefore,
one should start with the Lagrangian given in Eq.~(\ref{L_WB}), and
not Eq.~(\ref{L'}), since the ``K\"ahler potential'' for ${\cal L}'_0$
does not satisfy the condition (\ref{K_RS}).  For more on anomaly
mediation in the five-dimensional supergravity framework, see \cite{LS}.
}
\begin{equation}
\Delta m_{1/2}\ =\ -\frac{g^2}{16\pi^2}\, (T_G-T_R)\,
\frac{P_i F^i}{P}~.
\end{equation}

The Lagrangian ${\cal L}'$ can be made quantum mechanically
equivalent to ${\cal L}$ by adding an anomaly-canceling counterterm
to the bare Lagrangian ${\cal L}'_0$.  Such a counterterm is not
necessary for
the consistency of the low energy theory, nor is the choice of
counterterm unique.  Nevertheless, if we add the following
counterterm to ${\cal L}'_0$,
\begin{equation}
\Delta {\cal L}'\ =\ \frac{g^2}{32\pi^2}\, (T_G - T_R)\,
\int d^2\Theta ~2{\cal E}~ \log P ~W^\alpha W_\alpha\ +
\  {\rm h.c.}\ ,
\label{L_CT}
\end{equation}
then ${\cal L}' + \Delta {\cal L}'$ is equivalent to ${\cal L}$,
and the gaugino mass reduces to (\ref{gauginomasseinstein}).

\section{Conclusions}

In this paper we have seen that one-loop anomaly-induced gaugino
masses have a natural interpretation in terms of the supersymmetrized
Weyl, K\"ahler and sigma-model anomalies of matter-coupled supergravity.
Following Cardoso and Ovrut, we presented the superspace expression for
the nonlocal 1PI terms induced by these anomalies.  We then
found an expression for the visible-sector gaugino mass in terms of
the gravitino mass and
the hidden-sector auxiliary fields $F^i$.  Our expression holds for
arbitrary expectation values.  It reduces to previous results when
the hidden-sector expectation values are much smaller than
$M_{\rm P}$, and in the
``sequestered-sector" scenario of Randall and Sundrum \cite{RS}.

In the last section of the paper, we pointed out that the phenomenology
of the anomaly-induced gaugino masses is complicated
by possible anomaly-canceling
terms in the effective Lagrangian.  These terms can arise
from massive string modes, or from heavy supermultiplets whose masses
stem from symmetry breaking.  The string-induced terms are present
if some exact symmetry of string theory acts by a K\"ahler transformation
on the low-energy supergravity theory.  Their form
depends on the details of string theory at the Planck scale.

We also pointed out that the two forms of the supergravity action --
one where the superpotential $P$ is present, and the other where the
superpotential is absorbed in the K\"ahler potential $K$ -- are, in
fact, quantum-mechanically inequivalent.  In this paper we took as
fundamental
the action with the explicit superpotential.  This
action has a geometrical interpretation, in the sense that the
superpotential is a section of a holomorphic line bundle over the
K\"ahler manifold \cite{wittenbagger}.  More formal aspects of
supergravity anomalies will be investigated in Ref.~\cite{BMP2}.

Finally,
let us close by connecting our derivation of the anomaly-induced
mass to the discussion presented in the introduction.  There,
we stated that the gaugino mass follows from the
logarithmically-divergent term in the 1PI effective action,
\begin{equation}
\label{anomalybox}
{1 \over 4} \int d^2 \theta\, \left( 1\ -\ {g^2 b_0
\over 16 \pi^2}\, \log {\Lambda^2 \over \smallbox}
\right)\, W^\alpha  W_\alpha \ +\ {\rm h.c.}
\end{equation}
where $b_0 = 3 T_G - T_R$.
This expression assumes that all hidden-sector expectation
values are
much smaller than the Planck scale, and that spacetime is
flat and Minkowski.  If we interpret the cutoff as a
supersymmetry-breaking spurion, $\Lambda \rightarrow
\Lambda(1+ \theta^2 m_{3/2})$, Eq.~(\ref{anomalybox}) gives
rise to the nonvanishing gaugino mass (\ref{mass}).

Suppose now that we choose another cutoff, perhaps one that
preserves supersymmetry.  How then can we find the anomaly-induced
gaugino mass?  The answer is to write (\ref{anomalybox}) in a weak
supergravity background.  In particular,
we take the background
to be superconformally flat, with $E_m{}^a = \eta_m{}^a \, e^{\Sigma
+ \Sigma^+}$, where $\Sigma$ is a $\theta$-dependent but $x$-independent
conformal factor.  We replace the flat-space Laplacian by
$\smallbox \,e^{-4 \Sigma}$, which is the relevant part of the
covariant, chiral, curved-space Laplacian in the superconformal
background \cite{1001}.  This gives
\begin{eqnarray}
&&{1 \over 4} \int d^2 \theta\, \left( 1\ -\ {g^2 b_0
\over 16 \pi^2}\, \left [\, \log \left({\Lambda^2 \over \smallbox}
e^{4 \Sigma} \right) \right]
\right)\, W^\alpha  W_\alpha \ +\ {\rm h.c.}~\nonumber\\
&=&{1 \over 4} \int d^2 \theta\, \left( 1\ -\ {g^2 b_0
\over 16 \pi^2}\, \left [\, \log \left({\Lambda^2 \over \smallbox}
\right)   + 4 \Sigma\,\right] \right)\, W^\alpha  W_\alpha \ +
\ {\rm h.c.}
\end{eqnarray}
In the superconformal background, $R = -{\bar D}^2 \Sigma^+/4 + \cdots$,
so this becomes
\begin{equation}
{1 \over 4} \int d^2 \theta\, \left( 1\ -\ {g^2 b_0
\over 16 \pi^2}\, \left[\, \log \left({\Lambda^2 \over \smallbox}
\right) -  {{\bar D}^2 \over\smallbox}\, R^+ \right] \right)\,
W^\alpha
W_\alpha \ + \ {\rm h.c.}~
\end{equation}
plus higher-order terms.  The term proportional to $R^+$ is nothing
but the Weyl anomaly term from  (\ref{ovrutterm}).

We would like to thank Jan Louis, Lisa Randall, Riccardo
Rattazzi, Raman Sundrum and Fabio Zwirner for discussions.
The work of J.B.\ is supported by the
U.S.\ National Science Foundation, grants NSF-PHY-9404057 and
NSF-PHY-9970781.  He would like to thank the Institute
for Advanced Study and the Monell Foundation for their
support during the early stages of this project.
The work of T.M.\ is supported by U.S.\ National Science
Foundation under grant NSF-PHY-9513835, and by the Marvin
L.~Goldberger Membership.
E.P.\ is supported by the Department of Energy, contract
DOE DE-FG0292ER-40704.  We would like to express our
appreciation to the Aspen Center for Physics, where
part of this work was done.

\end{document}